# A Novel Policy Iteration Algorithm for Nonlinear Continuous-Time $H_\infty$ Control Problem

Qi Wang*

*China Airborne Missile Academy, Luoyang, Henan, 471009, People's Republic of China*

## I. Introduction

THE optimal control and optimal $H_\infty$ control have been the focus of many experts for many years. From the dynamic programming principle, the solution of optimal control problems or $H_\infty$ optimal control problems can be pursued by solving Hamilton–Jacobi–Bellman (HJB) equation or Hamilton–Jacobi–Isaacs (HJI) equation respectively [1], [2]. The HJB and HJI equations are formulated using performance value function, once the value function is solved, a feedback controller which results in closed-loop systems asymptotic stability can be synthesized. For linear systems and quadratic performance functionals, the HJB and HJI equations become the well-known algebraic Riccati equation and game algebraic Riccati equation [3]-[5], respectively. But for nonlinear systems or nonquadratic performance functionals, due to the inherent nonlinear properties, the solutions of HJB and HJI equation are challenging. Therefore, it is a formidable task to compute the value function of nonlinear systems by solving the HJB or HJI equation. Moreover, another disadvantage of solving HJB or HJI equation is that the full knowledge of the system dynamics needs to be available.

A great deal of effort has been made to solve these equations. Because it is intractable to get the accurate analytical solutions of HJB and HJI equations, approximate solutions of HJB and HJI have been found using many techniques, such as Galerkin's approximation of solving generalized HJB equation [6], adaptive dynamic programming [7], intelligent optimal control [8] and Taylor series solution [9]. For nonlinear continuous-time systems with control input constraint, an offline neural-networks-based HJB technique was proposed in [10], this method requires the full knowledge of systems model. To relax the requirement of internal dynamics knowledge of the nonlinear system, an online direct adaptive optimal control algorithm was presented in [11]. The experience replay technique was subsequently used to learn the solutions of HJB equation online for constrained-input systems with unknown internal dynamics in [12]. To overcome the difficulties of full knowledge requirements of the system model, a data-driven model-free approximate policy iteration (PI) algorithm was introduced in [13], [14], in which the off-policy learning method of reinforcement learning (RL) was used.

* Senior Engineer, Department of General System Technology, wangqibuaa@126.com.

The optimal $H_\infty$ control theory of nonlinear system was developed in [2], in which the notion of dissipativity introduced in [15] was utilized, and Schaft proofed that the smooth solution of the HJI equation can be pursued by a sequence of iterative procedures. Several offline algorithms of policy iteration for solving HJI equation were proposed in [16]-[19], in which solving the HJI equation was converted into solving a series of linear partial differential equation. In the offline policy iteration approach, the control input and disturbance input were computed iteratively in the inner-loop and outer-loop, respectively. Vamvoudakis *et al*. [20] introduced an online synchronous policy iteration algorithm to solve HJI equation, and Modares *et al*. [21] extended this method to constrained input systems. These methods need to tune synchronously three NN approximators, i.e., actor, disturbance, and critic, for calculating the approximate solution of the HJI equation. In [22]-[24], Luo *et al*. presented a simultaneous policy update algorithm. There need only one critic neural network to approximate the value function and only one iterative loop, thus, the simultaneous policy update algorithm is easy to offline or online implement. But the simultaneous policy update algorithm is an on-policy integral reinforcement learning algorithm, sampling data used in the policy evaluation phase are generated from evaluating policies. It requires that the disturbance input is adjustable, which is generally unrealistic for most practical systems. And then, an off-policy integral RL approach was introduced for $H_\infty$ control of nonlinear continuous-time system without requiring knowledge of the system internal dynamics [25]. The sampling data which are used for learning value function are generated with behavior policies rather than the target policies to be evaluated in the off-policy integral RL approach. A model-free version of off-policy integral RL approach for $H_\infty$ optimal control was proposed in [26]. Modares and Kiumarsi *et al*. extended the results of the simultaneous policy update algorithm and off-policy integral RL algorithm to the $H_\infty$ tracking control for nonlinear affine and nonaffine systems without using any knowledge of the dynamics of the systems [27], [28]. The algorithms for solving optimal control and $H_\infty$ control problems mentioned above are essentially Newton's iterative algorithms when a fixed-point equation is constructed in a Banach space. Hence, the simultaneous policy update algorithm and off-policy integral RL have the same disadvantage as the Newton iterative approach, that is, the algorithm is locally optimized, and finding suitable initial parameters or developing global methods is a difficult problem.

In recent years, Luo *et al*. presented an integration scheme of the policy iteration and value iteration methods to relax the requirement of the initial admissible control input and speed up the training convergence process for nonlinear discrete-time systems [29]. Besides, a model-free λ-PI algorithm was presented in [30], which can balance the control performance and convergence rate between PI algorithm and value iteration algorithm for linear discrete-time systems. Till present, the development of policy iteration algorithms and theories for $H_\infty$ control design of continuous-time nonlinear system is still an open issue, which promotes this research. It is worth noting that damped Newton's method, which leverages a non-unit step-size in the Newton direction by multiplying a damping parameter α, is an effective algorithm for solving equations. Although damped Newton method makes the convergence slower than Newton method, damped Newton's method could be globally convergent. Therefore, the damped Newton's method has the advantage of enhancing the robustness of the convergence of solving process with regarding to the initial guess.

In this paper, the $H_\infty$ control design of nonlinear continuous-time systems is considered, and a novel policy iteration algorithm based on damped Newton's method is developed for solving HJI equation, named α-policy iteration (α-PI) algorithm. By constructing a damped Newton iteration operator equation, a generalized Bellman equation is given, the generalized Bellman equation is an extension of Bellman equation. And then, by iterating on the generalized Bellman equation, the optimal solution of the HJI equation can be obtained. Under this framework, firstly, an on-policy α-PI integral reinforcement learning method is given, and then the off-policy α-PI reinforcement learning method is provided, namely a model-free method.

The organization of this paper is as follows. The continuous-time nonlinear $H_\infty$ control problem statement and preliminaries are presented in section II. The damped Newton method based on-policy and off-policy α-PI integral RL methods are provided in section III and IV respectively, and neural-network based implementation schemes of these α-PI algorithms are derived. Two computer simulation examples are presented in section V, one is for the general nonlinear system $H_\infty$ control problem, and the other is missile interception simulation. Finally, section VI concludes the paper.

## II. $H_\infty$ Control Problem and Simultaneous Policy Update Algorithm

In this section, the background knowledge reviews and preliminary results of $H_\infty$ control problem are provided. Consider a class affine in control input nonlinear continuous-time system defined as follows:

$$\dot{x} = f(x) + g(x)u + k(x)w \qquad (1)$$
$$z = h(x) \qquad (2)$$

where $x \in \Omega \subset \Re^n$ is state vector of the system, and $w \in \Re^q$ and $u \in \Re^m$ represent disturbance input vector and control input vector respectively, and $z \in \Re^p$ is the output of system (1). $f(x) \in \Re^n$ is internal dynamics, $g(x) \in \Re^{n \times m}$ and $k(x) \in \Re^{n \times q}$ are input-to-state matrix and disturbance coefficient matrix, respectively. On a compact set $\Omega$, $f(x)$, $g(x)$ and $k(x)$ meet locally Lipschitz continuity and $f(0) = 0$, that is $x = 0$ is an equilibrium point of the system. Assume that $w(t) \in L_2[0, \infty)$ and $u(t) \in L_2[0, \infty)$.

The objective of $H_\infty$ control problem is to reduce the influence of disturbance signal $w$ on the system, i.e., find a feedback control law $u$ such that system (1) meets a disturbance attenuation condition with starting from $x(0) = 0$, as follows:

$$\int_0^\infty (Q(x) + u^T R u)\, dt \leq \int_0^\infty \gamma^2 w^T w\, dt \tag{3}$$

where $Q(x) = \|z\|^2$, $R \in \Re^{m \times m}$ a positive definite symmetric matrix, and $\gamma > 0$ is a predetermined disturbance attenuation level.

If the condition (3) is satisfied, then the $L_2$-gain of system (1) is less than or equal to $\gamma > 0$, that is,

$$\frac{\int_0^\infty (Q(x) + u^T R u)\, dt}{\int_0^\infty w^T w\, dt} \leq \gamma^2. \tag{4}$$

The $H_\infty$ optimal control problem is concerned with finding the minimum $\gamma^*$ in the presence of solutions for bounded $L_2$-gain control problem. However, for general nonlinear system, there is no way to find the minimum value of the disturbance attenuation. Therefore, this paper focuses on suboptimal $H_\infty$ control problem, it is assumed that $\gamma$ is prescribed and $\gamma > \gamma^*$. It is worth noting that the two-person zero-sum differential games problem is equivalent to the suboptimal $H_\infty$ control problem [31]. Game theory deals with the strategic interaction between the two participants, each of which has an objective contained in a performance value function that the two participants try to maximize and minimize respectively.

Define the performance cost function for the policies of the two players as follows:

$$V^{u,w}(x(t)) = \int_t^\infty (Q(x) + u^T R u - \gamma^2 w^T w)\, dt. \tag{5}$$

In two-person zero-sum differential games, the objective of participant $u$ is to minimize the performance cost function, whereas the objective of participant $w$ is to maximize it, hence,

$$V^*(x_0) = \min_u \max_w \int_0^\infty (Q(x) + u^T R u - \gamma^2 w^T w)\, dt. \tag{6}$$

Define the Hamiltonian function related to the minimax design problem as

$$H(x,u,w,\nabla V) = Q(x) + u^T R u - \gamma^2 w^T w + \nabla V^T (f(x)+g(x)u+k(x)w) \tag{7}$$

where $\nabla V = \frac{\partial V}{\partial x}$.

In the sight of the Bellman optimality principle, if there exists a nonnegative solution $V^*(x) \geq 0: \Re^n \to \Re$ of HJI equation shown as follows:

$$0 = Q(x) + \left(\frac{\partial V^*}{\partial x}\right)^T f(x) - \frac{1}{4}\left(\frac{\partial V^*}{\partial x}\right)^T gR^{-1}g^T \frac{\partial V^*}{\partial x} + \frac{1}{4\gamma^2}\left(\frac{\partial V^*}{\partial x}\right)^T kk^T \frac{\partial V^*}{\partial x}. \tag{8}$$

Then, the control strategy pairs given in (9) and (10) are Nash equilibrium saddle-point solution [21].

$$u^* = -\frac{1}{2} R^{-1} g^T \frac{\partial V^*}{\partial x} \tag{9}$$

$$w^* = \frac{1}{2\gamma^2} k^T \frac{\partial V^*}{\partial x} \tag{10}$$

The Nash equilibrium strategy is inherently robust, once the equilibrium is reached, no participant can unilaterally deviate from its Nash strategy to improve its payoff. And, the $L_2$-gain of closed-loop system (1) with feedback control law (9) is less than or equal to $\gamma$.

Because the HJI equation is nonlinear with respect to $V^*$, finding the closed-form solution is usually intractable. The Bellman equation (11) which is linear in the cost function is used in simultaneous policy update algorithm for pursuing the solution of the HJI equation. Algorithm 1 shows the simultaneous policy update algorithm which is used to solve iteratively the $H_\infty$ control problem [24].

---

**Algorithm 1 Simultaneous Policy Update Algorithm for $H_\infty$ Control Problem**

Step 1: Set $i = 0$, give an initial value function $V_0$, and initial disturbance and control policies associated.

$$u_0 = -\frac{1}{2} R^{-1} g^T \frac{\partial V_i}{\partial x}, w_0 = \frac{1}{2\gamma^2} k^T \frac{\partial V_i}{\partial x}.$$

Step 2: Solve for $V_{i+1}(x)$ with $V_{i+1}(0) = 0$ by using

$$0 = Q(x) + u_i^T R u_i - \gamma^2 w_i^T w_i + \nabla V_{i+1}^T (f(x)+g(x)u_i+k(x)w_i). \tag{11}$$

Step 3: Update the disturbance policy and control policy using

$$u_{i+1} = -\frac{1}{2} R^{-1} g^T \frac{\partial V_{i+1}}{\partial x} \tag{12}$$

$$w_{i+1} = \frac{1}{2\gamma^2} k^T \frac{\partial V_{i+1}}{\partial x}. \tag{13}$$

Step 4: If $\|V_{i+1}-V_i\|_\Omega \leq \varepsilon$, $\varepsilon > 0$ (positive real number and small), stop calculation, else set $i = i+1$, move to Step 2 and go on iteration.

Luo *et al.* proofed that simultaneous policy update algorithm is essentially a Newton iteration algorithm for pursuing a solution of the fixed-point equation in a Banach space, The convergence can be established by the Kantorovich's theorem [23]. The proof method is briefly described as follows.

Firstly, considering a Banach space $\mathbb{V}$, $\mathbb{V} = \{V(x)|V(x): \Omega \to \mathfrak{R}, V(0)=0\}$, and $\|V\|$ is the norm of $V$. Define the following mapping $\mathcal{G}: \mathbb{V} \to \mathbb{V}$ on $\mathbb{V}$:

$$\mathcal{G}(V) = Q(x) + \left(\frac{\partial V}{\partial x}\right)^{\mathrm{T}} f(x) - \frac{1}{4}\left(\frac{\partial V}{\partial x}\right)^{\mathrm{T}} gR^{-1}g^{\mathrm{T}}\frac{\partial V}{\partial x} + \frac{1}{4\gamma^2}\left(\frac{\partial V}{\partial x}\right)^{\mathrm{T}} kk^{\mathrm{T}}\frac{\partial V}{\partial x} \tag{14}$$

The Fréchet derivative of mapping $\mathcal{G}(V)$ is denoted as $\mathcal{G}'(V)$, and the Fréchet differential is written as $\mathcal{G}'(V)Z$, where $Z \in \widetilde{\mathbb{V}} \subset \mathbb{V}$ and $\widetilde{\mathbb{V}}$ is a neighborhood of $V$.

Reference [23] showed that how to compute the Fréchet differential, as shown in Lemma 1.

*Lemma 1 ([23, Lemma 3])*: On the Banach space $\mathbb{V}$, $\forall V \in \mathbb{V}$, consider the mapping $\mathcal{G}$ defined in (14), the Fréchet differential of $\mathcal{G}$ is computed as:

$$\mathcal{G}'(V)Z = (\nabla Z)^{\mathrm{T}} f(x) - \frac{1}{4}(\nabla Z)^{\mathrm{T}} gR^{-1}g^{\mathrm{T}}\nabla V - \frac{1}{4}(\nabla V)^{\mathrm{T}} gR^{-1}g^{\mathrm{T}}\nabla Z + \frac{1}{4\gamma^2}(\nabla V)^{\mathrm{T}} kk^{\mathrm{T}}\nabla Z + \frac{1}{4\gamma^2}(\nabla Z)^{\mathrm{T}} kk^{\mathrm{T}}\nabla V \tag{15}$$

Then, in Theorem 1 of [23], Luo *et al.* proofed that the simultaneous policy update algorithm is equivalent mathematically to Newton method.

*Lemma 2 ([23, Theorem 1])*: The iteration procedure from (11) to (13) is equivalent to Newton method with (12) and (13), as follows:

$$V_{i+1} = V_i - \left(\mathcal{G}'(V_i)\right)^{-1}\mathcal{G}(V_i), \quad i = 0, 1, 2, \cdots. \tag{16}$$

In a Banach space, under some proper assumptions, the convergence of Newton method can be guaranteed by the Kantorovich's theorem [32]. And the solution of $\mathcal{G}(V^*) = 0$ can be obtained, thus, HJI equation (8) is solved.

## III. On-Policy $\alpha$-PI for Learning the Solution of HJI Equation

In this section, an on-policy $\alpha$-PI algorithm which is based on damped Newton iteration method is presented to pursue the solution of the HJI equation, and the $H_\infty$ controller is obtained. Then, NN-based critic–actor–disturbance scheme is leveraged to implement the on-policy $\alpha$-PI algorithm proposed without making use of the information about the internal dynamics of the system.

## A. On-Policy $\alpha$-PI Algorithm

Damped Newton's method is an effective approach for solving nonlinear equation, in which a non-unit step size in the Newton direction is leveraged by introducing a damping parameter. The major advantage of the damped Newton's method is that the convergence of solving process with regarding to the initial condition is enhanced.

Based on damped Newton method, the on-policy $\alpha$-PI is given in Algorithm 2.

---
**Algorithm 2 On-Policy $\alpha$-PI for $H_\infty$ Control Design**

Step 1: Set $i = 0$, give an initial value function $V_0$ with $V_0(0) = 0$, associated control and disturbance policies
$\boldsymbol{u}_0 = -\frac{1}{2}\boldsymbol{R}^{-1}\boldsymbol{g}^\text{T}\frac{\partial V_0}{\partial \boldsymbol{x}}, \boldsymbol{w}_0 = \frac{1}{2\gamma^2}\boldsymbol{k}^\text{T}\frac{\partial V_0}{\partial \boldsymbol{x}}$.

Step 2: Solve for $V_{i+1}(\boldsymbol{x})$ with $V_{i+1}(0) = 0$ by using

$$[V_{i+1}(\boldsymbol{x}(t+\Delta t))-V_{i+1}(\boldsymbol{x}(t))] + \alpha \int_t^{t+\Delta t}\left(Q(\boldsymbol{x})+\boldsymbol{u}_i^\text{T}\boldsymbol{R}\boldsymbol{u}_i-\gamma^2\boldsymbol{w}_i^\text{T}\boldsymbol{w}_i\right)\mathrm{d}t$$
$$= (1-\alpha)[V_i(\boldsymbol{x}(t+\Delta t))-V_i(\boldsymbol{x}(t))]. \qquad (17)$$

Step 3: Update the disturbance and control policies using

$$\boldsymbol{u}_{i+1} = -\frac{1}{2}\boldsymbol{R}^{-1}\boldsymbol{g}^\text{T}\frac{\partial V_{i+1}}{\partial \boldsymbol{x}} \qquad (18)$$

$$\boldsymbol{w}_{i+1} = \frac{1}{2\gamma^2}\boldsymbol{k}^\text{T}\frac{\partial V_{i+1}}{\partial \boldsymbol{x}}. \qquad (19)$$

Step 4: If $\|V_{i+1}-V_i\|_\Omega \leq \varepsilon$, $\varepsilon > 0$ (small positive number), stop calculation, else set $i = i + 1$, move to Step 2 and go on iteration.

---

In a Banach space $\mathbb{V}$, the above on-policy $\alpha$-PI algorithm is equivalent to damped Newton iteration method mathematically.

Considering the Banach space mentioned above, $\mathbb{V} = \{V(\boldsymbol{x})|V(\boldsymbol{x}): \Omega \to \mathfrak{R}, V(0)=0\}$, a new mapping $\mathcal{T}: \mathbb{V} \to \mathbb{V}$ is defined as follows:

$$\mathcal{T}(V) = V - \alpha\bigl(\mathcal{G}'(V)\bigr)^{-1}\mathcal{G}(V) \qquad (20)$$

where $\alpha$ is Newton step-size, $0 < \alpha \leq 1$.

*Theorem 1*: Given a mapping $\mathcal{T}: \mathbb{V} \to \mathbb{V}$ shown in (20), then the following damped Newton iteration with (18) and (19) is equivalent to the on-policy $\alpha$-PI algorithm from (17) to (19).

$$V_{i+1} = \mathcal{T}(V_i) \qquad (21)$$

*Proof*: Substituting (20) into (21), one can obtain the following equation

$$V_{i+1} = \mathcal{T}(V_i) = V_i - \alpha\big(\mathcal{G}'(V_i)\big)^{-1}\mathcal{G}(V_i). \tag{22}$$

Equation (22) can be rewritten as

$$\mathcal{G}'(V_i)V_{i+1} = \mathcal{G}'(V_i)V_i - \alpha\mathcal{G}(V_i). \tag{23}$$

From (18), (19), and (15), one has

$$\mathcal{G}'(V_i)V_{i+1} = (\nabla V_{i+1})^\mathrm{T}\boldsymbol{f}(\boldsymbol{x}) - \frac{1}{4}(\nabla V_{i+1})^\mathrm{T}\boldsymbol{g}\boldsymbol{R}^{-1}\boldsymbol{g}^\mathrm{T}\nabla V_i - \frac{1}{4}(\nabla V_i)^\mathrm{T}\boldsymbol{g}\boldsymbol{R}^{-1}\boldsymbol{g}^\mathrm{T}\nabla V_{i+1} + \frac{1}{4\gamma^2}(\nabla V_i)^\mathrm{T}\boldsymbol{k}\boldsymbol{k}^\mathrm{T}\nabla V_{i+1}$$

$$+ \frac{1}{4\gamma^2}(\nabla V_{i+1})^\mathrm{T}\boldsymbol{k}\boldsymbol{k}^\mathrm{T}\nabla V_i$$

$$= (\nabla V_{i+1})^\mathrm{T}\left[\boldsymbol{f}(\boldsymbol{x}) - \frac{1}{2}\boldsymbol{g}\boldsymbol{R}^{-1}\boldsymbol{g}^\mathrm{T}\nabla V_i + \frac{1}{2\gamma^2}\boldsymbol{k}\boldsymbol{k}^\mathrm{T}\nabla V_i\right]$$

$$= (\nabla V_{i+1})^\mathrm{T}[\boldsymbol{f}(\boldsymbol{x}) + \boldsymbol{g}\boldsymbol{u}_i + \boldsymbol{k}\boldsymbol{w}_i] \tag{24}$$

and

$$\mathcal{G}'(V_i)V_i = (\nabla V_i)^\mathrm{T}\boldsymbol{f}(\boldsymbol{x}) - \frac{1}{4}(\nabla V_i)^\mathrm{T}\boldsymbol{g}\boldsymbol{R}^{-1}\boldsymbol{g}^\mathrm{T}\nabla V_i - \frac{1}{4}(\nabla V_i)^\mathrm{T}\boldsymbol{g}\boldsymbol{R}^{-1}\boldsymbol{g}^\mathrm{T}\nabla V_i + \frac{1}{4\gamma^2}(\nabla V_i)^\mathrm{T}\boldsymbol{k}\boldsymbol{k}^\mathrm{T}\nabla V_i$$

$$+ \frac{1}{4\gamma^2}(\nabla V_i)^\mathrm{T}\boldsymbol{k}\boldsymbol{k}^\mathrm{T}\nabla V_i$$

$$= (\nabla V_i)^\mathrm{T}\boldsymbol{f}(\boldsymbol{x}) - 2\boldsymbol{u}_i^\mathrm{T}\boldsymbol{R}\boldsymbol{u}_i + 2\gamma^2\boldsymbol{w}_i^\mathrm{T}\boldsymbol{w}_i. \tag{25}$$

Substitute (24), (25) and (14) into (23), and make some mathematical manipulate, one can obtain the generalized Bellman equation, as follows:

$$(\nabla V_{i+1})^\mathrm{T}[\boldsymbol{f}(\boldsymbol{x}) + \boldsymbol{g}\boldsymbol{u}_i + \boldsymbol{k}\boldsymbol{w}_i] = (1-\alpha)\nabla V_i^\mathrm{T}[\boldsymbol{f}(\boldsymbol{x}) + \boldsymbol{g}\boldsymbol{u}_i + \boldsymbol{k}\boldsymbol{w}_i] - \alpha\big(Q(\boldsymbol{x}) + \boldsymbol{u}_i^\mathrm{T}\boldsymbol{R}\boldsymbol{u}_i - \gamma^2\boldsymbol{w}_i^\mathrm{T}\boldsymbol{w}_i\big). \tag{26}$$

Along the trajectory of system $\dot{\boldsymbol{x}} = \boldsymbol{f}(\boldsymbol{x}) + \boldsymbol{g}\boldsymbol{u}_i + \boldsymbol{k}\boldsymbol{w}_i$, integrating both sides of (26) from $t$ to $t + \Delta t$ yields the discretized version of generalized Bellman equation (17). In line with [3] and [25], with $V_{i+1}(0) = 0$, the differential version of generalized Bellman equation (26) and the discretized version of generalized Bellman equation (17) have the same solution.

This means that the damped Newton iteration given by (21) with (18) and (19) is equivalent to the on-policy $\alpha$-PI algorithm from (17) to (19). This completes the proof. □

*Remark 1:* The generalized Bellman equation (26) is linear partial differential equation just like the Bellman equation (11).

*Remark 2*: Compared with Bellman equation (11), the integral reinforcement term of the generalized Bellman equation (26) is multiplied by the damped coefficient $\alpha$, and reinforcement term needs to consider the influence of the current value function $V_i$.

*Remark 3*: When $\alpha = 1$, (26) degenerates to Bellman equation (11), and on-Policy $\alpha$-PI algorithm becomes the simultaneous policy update algorithm shown in Algorithm 1.

Newton method has a faster convergence rate near the solution, but smaller convergence region, it usually only satisfies local convergence and needs a good initial guess. On the other hand, damped Newton method converges slowly, but convergence condition can be relaxed. By choosing $\alpha$ smaller one can obtain larger convergence domain [33].

In order to obtain the optimal control policy, the value function $V_{i+1}$ needs to be solved in algorithm 2. Due to the unknown form of the value function, a NN approximator is used to approximate the value function. A NN-based implementation scheme of the iteration algorithm 2 is provided in the next section.

**B. NN-Based Online Implement of Algorithm 2**

Because of the universal approximation property, NNs are natural candidates for approximating smooth functions on compact sets. In this section, we use a single critic NN structure to implement on-policy $\alpha$-PI algorithm online.

Let $\boldsymbol{\rho}(x) = [\rho_1(x), \cdots, \rho_N(x)]^{\mathrm{T}}$ be the critic NN activation functions, it means that there are $N$ neurons on the hidden layer. The cost function is approximated by

$$V(x) = \hat{V}(x) + \varepsilon(x) = \boldsymbol{W}^{\mathrm{T}} \boldsymbol{\rho}(x) + \varepsilon(x) \tag{27}$$

where $\boldsymbol{W} \in \mathfrak{R}^N$ is the weight vector of NN, $\varepsilon(x)$ is the error of NN approximator. Based on high-order Weierstrass approximation theories [34], if $V(x)$ is smooth enough, there exists a complete basis set, that $\varepsilon \to 0$ uniformly with $N \to \infty$.

By using NN approximator $\hat{V}_{i+1}(x) = \boldsymbol{W}_{i+1}^{\mathrm{T}} \boldsymbol{\rho}(x)$ and $\hat{V}_i(x) = \boldsymbol{W}_i^{\mathrm{T}} \boldsymbol{\rho}(x)$, given control $\boldsymbol{u}_i$ and disturbance $\boldsymbol{w}_i$, generalized Bellman equation (17) becomes

$$[\boldsymbol{W}_{i+1}^{\mathrm{T}} \boldsymbol{\rho}(x(t+\Delta t)) - \boldsymbol{W}_{i+1}^{\mathrm{T}} \boldsymbol{\rho}(x(t))] + \alpha \int_t^{t+\Delta t} (Q(x) + \boldsymbol{u}_i^{\mathrm{T}} \boldsymbol{R} \boldsymbol{u}_i - \gamma^2 \boldsymbol{w}_i^{\mathrm{T}} \boldsymbol{w}_i) \, \mathrm{d}t$$
$$= (1-\alpha)[\boldsymbol{W}_i^{\mathrm{T}} \boldsymbol{\rho}(x(t+\Delta t)) - \boldsymbol{W}_i^{\mathrm{T}} \boldsymbol{\rho}(x(t))]. \tag{28}$$

That is,

$$[\boldsymbol{\rho}^{\mathrm{T}}(x(t)) - \boldsymbol{\rho}^{\mathrm{T}}(x(t+\Delta t))] \boldsymbol{W}_{i+1} = (1-\alpha)[\boldsymbol{\rho}^{\mathrm{T}}(x(t)) - \boldsymbol{\rho}^{\mathrm{T}}(x(t+\Delta t))] \boldsymbol{W}_i$$
$$+ \alpha \int_t^{t+\Delta t} (Q(x) + \boldsymbol{u}_i^{\mathrm{T}} \boldsymbol{R} \boldsymbol{u}_i - \gamma^2 \boldsymbol{w}_i^{\mathrm{T}} \boldsymbol{w}_i) \, \mathrm{d}t. \tag{29}$$

It is worth noting that there are $N$ unknown parameters in the NN approximator. Therefore, at least $N$ sampling data points are needed to solve the weight vector of neural network while the batch least-squares (BLS) method is used. When $\bar{N}$ ($\bar{N} > N$) data are collected along the system state trajectories in the domain of $\Omega$, the BLS solution of the NN weights is

$$W_{i+1} = (XX^{\mathrm{T}})^{-1}XY \tag{30}$$

where

$$X = [\rho(x(t)) - \rho(x(t+\Delta t)), \cdots, \quad \rho(x(t+(\bar{N}-1)\Delta t)) - \rho(x(t+\bar{N}\Delta t))] \tag{31}$$

$$Y = [y^{u_i, w_i}(x(t)), \cdots, y^{u_i, w_i}(x(t+(\bar{N}-1)\Delta t))]^{\mathrm{T}} \tag{32}$$

and

$$y^{u_i, w_i}(x(t+k\Delta t)) = (1-\alpha)[\rho^{\mathrm{T}}(x(t+k\Delta t)) - \rho^{\mathrm{T}}(x(t+(k+1)\Delta t))]W_i$$

$$+ \alpha \int_{t+k\Delta t}^{t+(k+1)\Delta t} (Q(x) + u_i^{\mathrm{T}} R u_i - \gamma^2 w_i^{\mathrm{T}} w_i) \, \mathrm{d}t \tag{33}$$

with $k = 0, 1, \cdots, \bar{N} - 1$.

Having solved for the approximate value function $\hat{V}_{i+1}(x)$ associated with the disturbance $w_i$ and control $u_i$, the disturbance and control policies update step can be executed as follows:

$$\hat{w}_{i+1} = \frac{1}{2\gamma^2} k^{\mathrm{T}} (\nabla \rho)^{\mathrm{T}} W_{i+1} \tag{34}$$

$$\hat{u}_{i+1} = -\frac{1}{2} R^{-1} g^{\mathrm{T}} (\nabla \rho)^{\mathrm{T}} W_{i+1} \tag{35}$$

where $\nabla \rho = \frac{\partial \rho}{\partial x}$, is the Jacobian matrix of $\rho(x)$.

Since Bellman equation (11) and generalized Bellman equation (26) are first order linear partial differential equation, utilizing some important theorems and properties that have been formed in the field of computational intelligence, the convergence proof of the presented BLS NN algorithm for implementing Algorithm 2 is almost the same as that in [10] and [24].

## IV. Off-Policy $\alpha$-PI for Learning the Solution of HJI Equation

For on-policy $\alpha$-PI algorithm, the disturbance policy needs to be adjustable, which is usually unrealistic for practical applications. To overcome this drawback, motivated by [25]-[28], an off-policy $\alpha$-PI algorithm is proposed to pursue the solution of HJI equation without using any prior information about the dynamics of the system. Then, a NN-based online implementation scheme is given.

## A. Off-Policy $\alpha$-PI Algorithm

Firstly, the system dynamics (1) can be rewritten as

$$\dot{x} = f(x) + g(x)u_i + k(x)w_i + g(x)(u-u_i) + k(x)(w-w_i) \tag{36}$$

where $u \in \Re^m$ and $w \in \Re^q$ are behavior policy and actual disturbance, respectively. $u_i \in \Re^m$ and $w_i \in \Re^q$ are strategies to be evaluated and updated. Let $V_{i+1}(x)$ be the solution of generalized Bellman equation (26), and differentiating $V_{i+1}(x)$ along with the system dynamics (36) yields

$$\dot{V}_{i+1} = (\nabla V_{i+1})^T(f+gu_i+kw_i) + (\nabla V_{i+1})^T g(u-u_i) + (\nabla V_{i+1})^T k(w-w_i). \tag{37}$$

Using (26) one can obtain

$$\dot{V}_{i+1} = -\alpha\bigl(Q(x)+u_i^T R u_i - \gamma^2 w_i^T w_i\bigr) + (1-\alpha)\nabla V_i^T[f(x)+gu_i+kw_i] + (\nabla V_{i+1})^T g(u-u_i) + (\nabla V_{i+1})^T k(w-w_i). \tag{38}$$

Substituting (36) into (38), one can obtain

$$\begin{aligned}
\dot{V}_{i+1} &= -\alpha\bigl(Q(x)+u_i^T R u_i - \gamma^2 w_i^T w_i\bigr) + (1-\alpha)\nabla V_i^T[\dot{x}-g(x)(u-u_i)-k(x)(w-w_i)] \\
&\quad + (\nabla V_{i+1})^T g(u-u_i) + (\nabla V_{i+1})^T k(w-w_i) \\
&= -\alpha\bigl(Q(x)+u_i^T R u_i - \gamma^2 w_i^T w_i\bigr) + (1-\alpha)\nabla V_i^T \dot{x} - (1-\alpha)\nabla V_i^T g(x)(u-u_i) - (1-\alpha)\nabla V_i^T k(x)(w-w_i) \\
&\quad + (\nabla V_{i+1})^T g(u-u_i) + (\nabla V_{i+1})^T k(w-w_i).
\end{aligned} \tag{39}$$

Substituting $u_{i+1} = -\frac{1}{2}R^{-1}g^T \frac{\partial V_{i+1}}{\partial x}$, $u_i = -\frac{1}{2}R^{-1}g^T \frac{\partial V_i}{\partial x}$, $w_{i+1} = \frac{1}{2\gamma^2}k^T \frac{\partial V_{i+1}}{\partial x}$ and $w_i = \frac{1}{2\gamma^2}k^T \frac{\partial V_i}{\partial x}$ into (39)

yields the following off-policy generalized Bellman equation:

$$\begin{aligned}
\dot{V}_{i+1} &= -\alpha\bigl(Q(x)+u_i^T R u_i - \gamma^2 w_i^T w_i\bigr) + (1-\alpha)\dot{V}_i + (1-\alpha)2u_i^T R(u-u_i) - (1-\alpha)2\gamma^2 w_i^T(w-w_i) \\
&\quad - 2u_{i+1}^T R(u-u_i) + 2\gamma^2 w_{i+1}^T(w-w_i).
\end{aligned} \tag{40}$$

It can be seen from the derivation process of (40) that it has the same solution as (26). Integrating both sides of (40) in interval $[t, t+\Delta t)$ forms the discretized version of off-policy generalized Bellman equation (41).

$$\begin{aligned}
&\bigl[V_{i+1}\bigl(x(t+\Delta t)\bigr)-V_{i+1}\bigl(x(t)\bigr)\bigr] + \int_t^{t+\Delta t} 2u_{i+1}^T R(u-u_i)\, dt - \int_t^{t+\Delta t} 2\gamma^2 w_{i+1}^T(w-w_i)\, dt \\
&= -\alpha \int_t^{t+\Delta t} \bigl(Q(x)+u_i^T R u_i - \gamma^2 w_i^T w_i\bigr)\, dt + (1-\alpha)\bigl[V_i\bigl(x(t+\Delta t)\bigr)-V_i\bigl(x(t)\bigr)\bigr] + (1-\alpha)\int_t^{t+\Delta t} 2u_i^T R(u-u_i)\, dt \\
&\quad - (1-\alpha)\int_t^{t+\Delta t} 2\gamma^2 w_i^T(w-w_i)\, dt
\end{aligned} \tag{41}$$

Equation (41) and generalized Bellman equation (26) have the same solution for the cost function. For space reasons the proof is omitted in this paper. It is observed from (41) that arbitrary input signals $u$ and $w$ can be used for learning the value function $V_{i+1}$, rather than the policies $u_i$ and $w_i$ to be evaluated. Then, replacing (17) in Algorithm 2 with (41), one can obtains the off-policy $\alpha$-PI Algorithm, as shown in Algorithm 3.

---

**Algorithm 3 Off-Policy $\alpha$-PI for $H_\infty$ Control Problem**

Step 1: Use the behavior policy $u$ and the actual disturbance $w$ to collect $M$ system data which contain system state, disturbance input and control input at different sampling time interval.

Step 2: Set $i = 0$, give an initial cost function $V_0$ with $V_0(0) = 0$, and initial control and disturbance policies
$u_0$ and $w_0$.

Step 3: Reuse the collected data to solve (41) for $V_{i+1}(x)$, $u_{i+1}$ and $w_{i+1}$, with $V_{i+1}(0) = 0$.

Step 4: If a stop iteration condition is met, stop iteration and output $V_{i+1}(x)$ as the approximate optimal solution of HJI equation (8), output $u_{i+1}$ as the approximate optimal control input, i.e., $V^*(x) = V_{i+1}(x)$ and $u^* = u_{i+1}$, else set $i = i + 1$, move to Step 3 and go on iteration.

---

Algorithm 3 is a model-free off-policy integral RL method, none of the prior information about system dynamics is needed. When $\alpha = 1$, (41) degenerates to off-policy integral reinforcement learning Bellman equation as shown in [26] and [27].

**B. NN-Based Online Implement of Algorithm 3**

In order to solve $V_{i+1}(x)$, $u_{i+1}$ and $w_{i+1}$ in (41) by using the system sampling data, the Neural Networks based actor-critic structure is introduced. Here three NNs, i.e., two actor NNs and one critic NN, are adopted to learn the cost function, disturbance policy and control policy, respectively. The three NNs are given as follows:

$$\hat{V}_{i+1}(x) = (W_c^{i+1})^\mathrm{T} \rho(x) \tag{42}$$

$$\hat{u}_{i+1}(x) = (W_a^{i+1})^\mathrm{T} \varphi(x) \tag{43}$$

$$\hat{w}_{i+1}(x) = (W_d^{i+1})^\mathrm{T} \phi(x) \tag{44}$$

where $\rho(x) = [\rho_1(x),...,\rho_{L_1}(x)]^\mathrm{T} \in \Re^{L_1}$ are the linearly independent basis functions for the critic NN, $\varphi(x) = [\varphi_1(x),...,\varphi_{L_2}(x)]^\mathrm{T} \in \Re^{L_2}$ and $\phi(x) = [\phi_1(x),...,\phi_{L_3}(x)]^\mathrm{T} \in \Re^{L_3}$ are the linearly independent basis functions for actor and disturbance NNs, respectively, which are all defined on $\Omega \subset \Re^n$. $L_1$, $L_2$ and $L_3$ are the number of neurons in the hidden layer of the three neural networks, respectively. $\boldsymbol{W}_c^{i+1} \in \Re^{L_1}$, $\boldsymbol{W}_a^{i+1} \in \Re^{L_2 \times m}$ and $\boldsymbol{W}_d^{i+1} \in \Re^{L_3 \times q}$ are weight vectors. Define $\boldsymbol{R} = \text{diag}(r_1,...,r_m)$, substituting (42)-(44) into (41) yields

$$(\boldsymbol{W}_c^{i+1})^{\text{T}}[\boldsymbol{\rho}(\boldsymbol{x}(t+\Delta t))-\boldsymbol{\rho}(\boldsymbol{x}(t))] + 2\sum_{j=1}^{m} r_j \int_t^{t+\Delta t} (\boldsymbol{W}_{a,j}^{i+1})^{\text{T}} \boldsymbol{\varphi}(\boldsymbol{x})\mu_j \, dt - 2\gamma^2 \sum_{k=1}^{q} \int_t^{t+\Delta t} (\boldsymbol{W}_{d,k}^{i+1})^{\text{T}} \boldsymbol{\phi}(\boldsymbol{x})\nu_k \, dt$$

$$= -\alpha \int_t^{t+\Delta t} \left(Q(\boldsymbol{x})+\boldsymbol{u}_i^{\text{T}}\boldsymbol{R}\boldsymbol{u}_i-\gamma^2 \boldsymbol{w}_i^{\text{T}}\boldsymbol{w}_i\right) dt + (1-\alpha)(\boldsymbol{W}_c^i)^{\text{T}}[\boldsymbol{\rho}(\boldsymbol{x}(t+\Delta t))-\boldsymbol{\rho}(\boldsymbol{x}(t))]$$

$$+ 2(1-\alpha)\sum_{j=1}^{m} r_j \int_t^{t+\Delta t} (\boldsymbol{W}_{a,j}^i)^{\text{T}} \boldsymbol{\varphi}(\boldsymbol{x})\mu_j \, dt - 2(1-\alpha)\gamma^2 \sum_{k=1}^{q} \int_t^{t+\Delta t} (\boldsymbol{W}_{d,k}^i)^{\text{T}} \boldsymbol{\phi}(\boldsymbol{x})\nu_k \, dt \quad (45)$$

where $\boldsymbol{\mu} = [\mu_1, \ldots, \mu_m]^{\text{T}} = \boldsymbol{u} - \boldsymbol{u}_i$, $\boldsymbol{v} = [\nu_1,...,\nu_q]^{\text{T}} = \boldsymbol{w} - \boldsymbol{w}_i$, $\boldsymbol{W}_{a,j}^{i+1}$ and $\boldsymbol{W}_{a,j}^i$ are the $j^{\text{th}}$ column of matrix $\boldsymbol{W}_a^{i+1}$ and $\boldsymbol{W}_a^i$ respectively, $\boldsymbol{W}_{d,k}^{i+1}$ and $\boldsymbol{W}_{d,k}^i$ are the $k^{\text{th}}$ column of matrix $\boldsymbol{W}_d^{i+1}$ and $\boldsymbol{W}_d^i$ respectively. Note that (45) is linear in the Neural Networks weight vectors $\boldsymbol{W}_c^{i+1}$, $\boldsymbol{W}_a^{i+1}$ and $\boldsymbol{W}_d^{i+1}$. Define

$$\boldsymbol{W}_1^{i+1} = \left[(\boldsymbol{W}_c^{i+1})^{\text{T}}, (\boldsymbol{W}_{a,1}^{i+1})^{\text{T}}, \ldots, (\boldsymbol{W}_{a,m}^{i+1})^{\text{T}}, (\boldsymbol{W}_{d,1}^{i+1})^{\text{T}}, \ldots, (\boldsymbol{W}_{d,q}^{i+1})^{\text{T}}\right]^{\text{T}} \quad (46)$$

$$\boldsymbol{\varpi}(t) = \begin{bmatrix} \boldsymbol{\rho}(\boldsymbol{x}(t+\Delta t)) - \boldsymbol{\rho}(\boldsymbol{x}(t)) \\ 2r_1 \int_t^{t+\Delta t} \boldsymbol{\varphi}(\boldsymbol{x})\mu_1 \, dt \\ \vdots \\ 2r_m \int_t^{t+\Delta t} \boldsymbol{\varphi}(\boldsymbol{x})\mu_m \, dt \\ -2\gamma^2 \int_t^{t+\Delta t} \boldsymbol{\phi}(\boldsymbol{x})\nu_1 \, dt \\ \vdots \\ -2\gamma^2 \int_t^{t+\Delta t} \boldsymbol{\phi}(\boldsymbol{x})\nu_q \, dt \end{bmatrix} \quad (47)$$

and

$$\lambda(t) = -\alpha \int_t^{t+\Delta t} \left(Q(\boldsymbol{x})+\boldsymbol{u}_i^{\text{T}}\boldsymbol{R}\boldsymbol{u}_i - \gamma^2 \boldsymbol{w}_i^{\text{T}}\boldsymbol{w}_i\right) dt + (1-\alpha)(\boldsymbol{W}_c^i)^{\text{T}}[\boldsymbol{\rho}(\boldsymbol{x}(t+\Delta t))-\boldsymbol{\rho}(\boldsymbol{x}(t))]$$

$$+2(1-\alpha)\sum_{j=1}^{m} r_j \int_t^{t+\Delta t} (\boldsymbol{W}_{a,j}^i)^{\text{T}} \boldsymbol{\varphi}(\boldsymbol{x})\mu_j \, dt - 2(1-\alpha)\gamma^2 \sum_{k=1}^{q} \int_t^{t+\Delta t} (\boldsymbol{W}_{d,k}^i)^{\text{T}} \boldsymbol{\phi}(\boldsymbol{x})\nu_k \, dt. \quad (48)$$

Then (45) can be rewritten as

$$\lambda(t) = (\boldsymbol{W}_1^{i+1})^{\text{T}} \boldsymbol{\varpi}(t). \quad (49)$$

Note that (49) is linear in the $\boldsymbol{W}_1^{i+1}$, therefore $\boldsymbol{W}_1^{i+1}$ can be solved in the sense of least-squares. Because of $\boldsymbol{W}_1^{i+1} \in \Re^{L_1+m \times L_2+q \times L_3}$, therefore one needs to collect $M > L_1 + m \times L_2 + q \times L_3$ system data about system state, disturbance and control input from $t_1$ to $t_M$ in the state space. Then, for the given evaluating policies $\widehat{w}_i$ and $\widehat{u}_i$, using this information to calculate (47) and (48) at the $M$ different sampling data point, one can get

$$\Pi = [\varpi(t_1),...,\varpi(t_M)] \tag{50}$$

$$\Lambda = [\lambda(t_1),...,\lambda(t_M)]^{\mathrm{T}}. \tag{51}$$

The least-squares solution of (49) is given as

$$W_1^{i+1} = (\Pi\Pi^{\mathrm{T}})^{-1}\Pi\Lambda. \tag{52}$$

Then one can obtain $\widehat{V}_{i+1}$, $\widehat{w}_{i+1}$ and $\widehat{u}_{i+1}$ using (42-44). Reusing the collect data to solve (49), one can get the approximate optimal control input pairs $u^*$ and $w^*$.

*Remark 4:* Although (47) contains $x(t+\Delta t)$, the data can be sampled continuously over the same time interval $\Delta t$, i.e., $x(t), x(t+\Delta t), ..., x(t+M\Delta t)$. Thus, none of the information about the dynamics of the system is needed for calculating the system state vector $x(t+\Delta t)$ at instant $t$.

## V. Simulation Validation

In this section, the effectiveness of the off-policy α-PI algorithm is verified through two computer simulation examples, one is $H_\infty$ control problem for the general nonlinear system, and the other is missile interception simulation.

### A. Nonlinear System $H_\infty$ Control Simulation

This is a revised version of example 2 in [23]. The model is as follows:

$$\dot{x} = \begin{bmatrix} -x_1 + x_2 \\ -0.5x_1 - 0.5x_2 + 0.5x_2\sin^2(x_1) \end{bmatrix} + \begin{bmatrix} 0 \\ \sin(x_1) \end{bmatrix} u + \begin{bmatrix} 0 \\ \cos(x_1) \end{bmatrix} w \tag{53}$$

$$z = x. \tag{54}$$

Select $R=1$ and $\gamma=2$ for performance value function (5). Polynomial functions are selected as the basis functions of neural network, that is, $\rho(x) = [x_1^2, x_2^2, x_1x_2, x_1^4, x_2^4]^{\mathrm{T}}$ with $L_1=5$, $\varphi(x) = [x_1, x_2, x_1^2, x_1x_2, x_2^2, x_1^2x_2, x_1x_2^2, x_2^3, x_1^3]^{\mathrm{T}}$ with $L_2=9$, and disturbance NN and control NN have the same basis functions, i.e., $\phi(x) = \varphi(x)$ with $L_3 = L_2$. Algorithm 3 is used to learn online the solution of HJI equation. During the data sampling phase, the initial state of the system is set as $x(0) = [0.4, 0.5]^{\mathrm{T}}$, sampling time step is $\Delta t = 0.05$ sec, and control input signal and disturbance input signal are set as random signal with uniform distributed in $[-1,1]$. And then collect 50 (i.e., $M=50$) sample data from $t = 0$ sec to $t = 2.5$ sec. At the learning stage, the initial weight vectors of the three NNs are set as zero, $W_c^0 = 0$, $W_a^0 = 0$ and $W_d^0 = 0$, it means that $V_0 = 0$, $u_0 = 0$ and $w_0 = 0$. The Newton step-size is set as $\alpha = 0.3$. By reusing the sampled data, the weight vector of NNs can be solved iteratively, when $\|W_1^{i+1} - W_1^i\| < 10^{-7}$ stop iteration.

Figs. 1 and 2 show the weight vector $\boldsymbol{W}_c$ of critic NN and the weight vector $\boldsymbol{W}_a$ of actor NN at each iteration. It can be seen that the weight vectors of NNs converge at $33^{rd}$ iteration, $\boldsymbol{W}_c^{33} = [0.5215, -0.0298, 1.1289, -0.0987, 0.0245]^T$, $\boldsymbol{W}_a^{33} = [\ -0.0056, 0.0157, 0.0719, -1.4079, 0.2066, 0.6638, -0.4448, -0.0721, -0.1117]^T$.

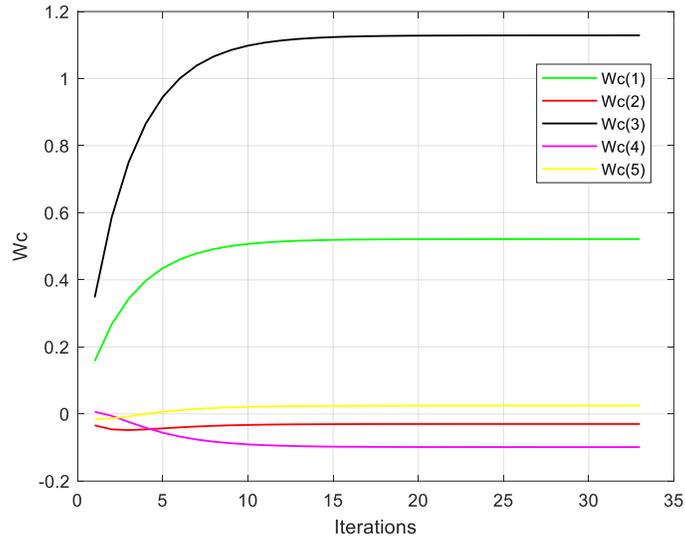

**Fig. 1 Weight vector $\boldsymbol{W}_c$ of critic NN.**

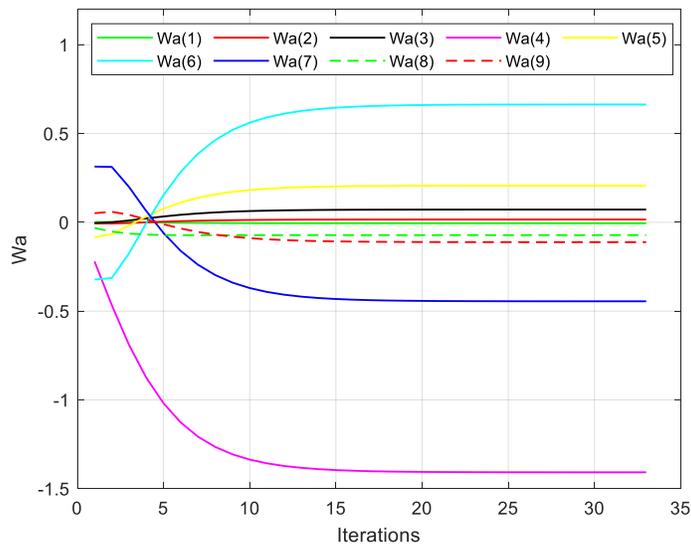

**Fig. 2 Weight vector $\boldsymbol{W}_a$ of actor NN.**

Leveraging the NNs weights learned, the solution of HJI equation and the corresponding $H_\infty$ controller can be obtained through (42) and (43). Restate the system with the controller obtained and set the disturbance signal as:

$$w = 5\exp(-0.2(t-t_0))\cos(t-t_0) \quad (55)$$

where $t_0 = 2.5$. The state trajectories of the closed-loop system, control input and disturbance signal are shown in Figs. 3 and 4. The first 2.5 seconds in Figs. 3 and 4 correspond to the data collected phase. At the instant 2.5 sec, off-policy α-PI is leveraged to pursue the solution of HJI equation. It also can be seen form Fig. 5 that the disturbance attenuation level is less than 0.6, which meets the $L_2$-gain condition.

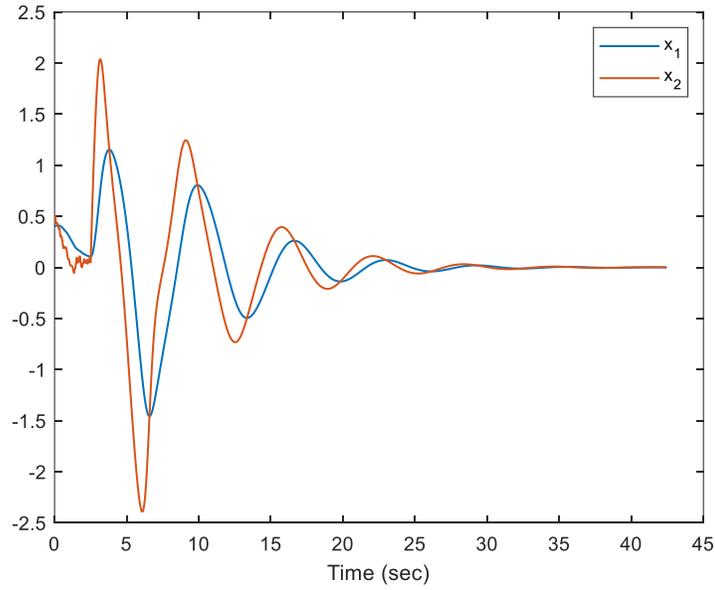

**Fig. 3 State trajectories of the closed-loop system.**

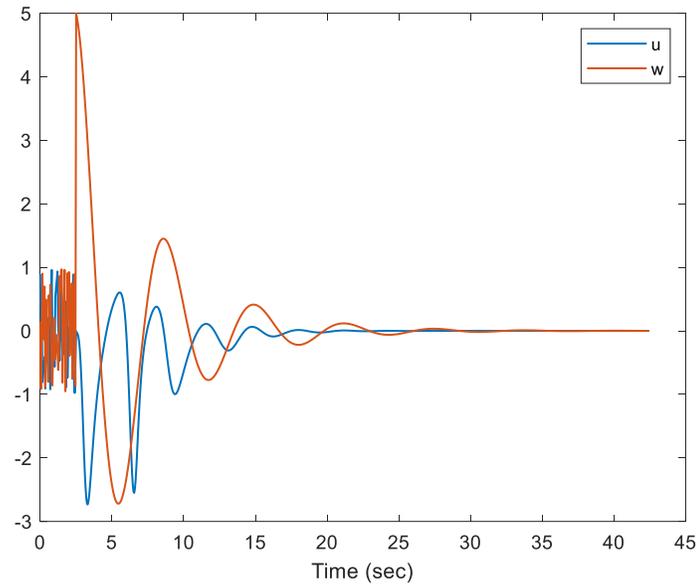

**Fig. 4  Control input and disturbance signals.**

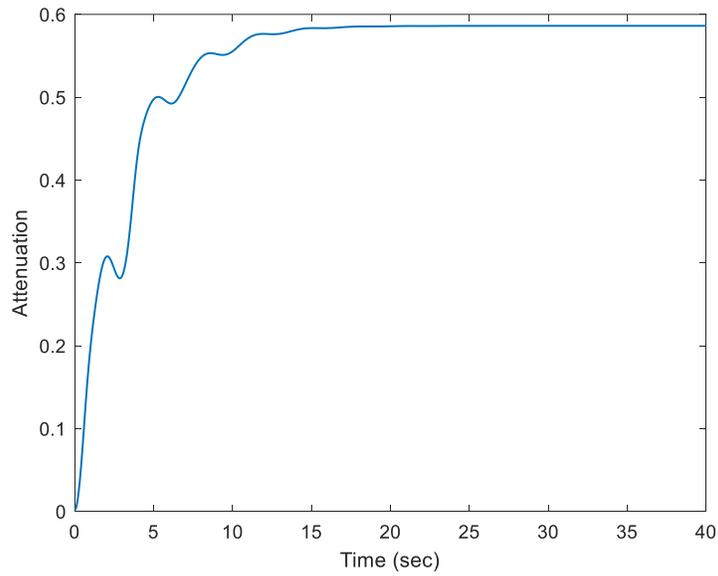

**Fig. 5  Attenuation level of disturbance with the controller obtained.**

## B. Missile Interception Simulation

The missile interception problem can be regarded as a two-person zero-sum differential games problem, which depends on the solution of HJI equation. The engagement geometry of missile interception problem is shown in Fig. 6.

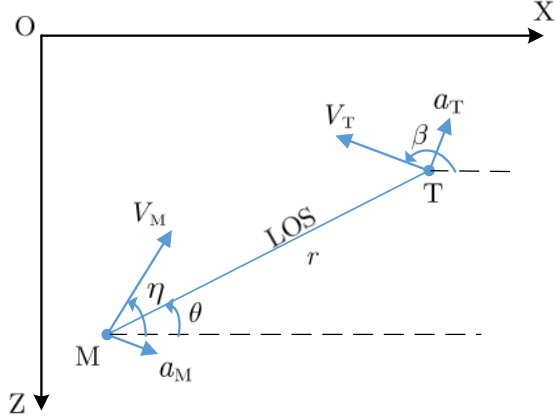

**Fig. 6 Missile interception target scenario.**

where $V_M$ represents missile velocity, $\eta$ is the flight-path angles (FPA) of missile, $a_M$ is missile acceleration which is perpendicular to its velocity. $V_T$, $\beta$ and $a_T$ are the corresponding variables of the target. $\theta$ and $r$ represent line-of-sight angle and missile-target relative distance, respectively. The relative motion equations of missile and target are as follows.

$$\dot{r} = V_r = V_T\cos(\beta - \theta) - V_M\cos(\eta - \theta) \tag{55}$$

$$\dot{\theta} = [V_T\sin(\beta - \theta) - V_M\sin(\eta - \theta)]/r \tag{56}$$

where $V_r$ is closing velocity, i.e., range rate along the line-of-sight, $\dot{\theta}$ is line-of-sight rate. Assume that the velocity of the missile and target is constant, select $x_1 = \theta$ and $x_2 = \dot{\theta}$ as the system state variables, take the derivative of both sides of (56), one can get

$$\begin{cases} \dot{x}_1 = x_2 \\ \dot{x}_2 = -\dfrac{2V_r}{r}x_2 + \dfrac{\cos(\eta - \theta)}{r}a_M - \dfrac{\cos(\beta - \theta)}{r}a_T \end{cases} \tag{57}$$

In the missile interception target problem, the expectation of missile is to minimize $\dot{\theta}$, however, the objective of target which is an opponent is to maximize $\dot{\theta}$. Therefore, the missile interception target problem can be regarded as a two-person zero-sum differential games problem [35].

The performance cost function for the policies of missile and target is defined as follows

$$V^{u,w}(x(t)) = \int_t^\infty (Q(x_1, x_2) + Ra_M^2 - \gamma^2 a_T^2)\, dt. \tag{58}$$

where $Q(x_1, x_2) = q_1 x_1^2 + q_2 x_2^2$, $q_1 = 0$, $q_2 = 10^8$, $R = 1$ and $\gamma = 10$.

The hidden layer neurons of critic NN, actor NN and disturber NN on $\Omega$ are set as follows, respectively.

$$\boldsymbol{\rho}(x_1, x_2) = [x_1^4, x_1^3 x_2, x_1^2 x_2^2, x_1 x_2^3, x_2^4, x_1^2, x_1 x_2, x_2^2]^T \in \mathbb{R}^8 \tag{59}$$

$$\boldsymbol{\phi}(x_1, x_2) = [x_2, x_1^2 x_2, x_2^3]^T \in \mathbb{R}^3 \tag{60}$$

$$\boldsymbol{\varphi}(x_1, x_2) = [x_2, x_1^2 x_2, x_2^3]^T \in \mathbb{R}^3 \tag{61}$$

Thus, $W_c \in \mathbb{R}^8$, $W_a \in \mathbb{R}^{3\times 1}$ and $W_d \in \mathbb{R}^{3\times 1}$, then $W \in \mathbb{R}^{14}$.

Next, the off-policy α-PI algorithm proposed in this paper is used to solve this missile interception problem. Because $r$, $V_r$, $\phi$ and $\beta$ change over time, (57) is a time-varying differential equation. Therefore, the off-policy α-PI algorithm is periodically investigated to solve the optimal control policy $a_M^*$ of the missile. The numerical simulation calculation step $\Delta t$ is select to be 0.005 s, and take $N = 100$, i.e., collect 100 data points every 0.5 s along the system state trajectories, therefore Algorithm 3 needs to be used for solving the solution of HJI equation every 0.5 s. The approximate optimal control strategy of the missile obtained in the previous cycle will be used as behavior policy of the missile in the next cycle.

Numerical simulation conditions of missile and target engagement scenario are shown in Table 1.

**Table 1 Numerical simulation conditions**

| Parameters | Symbol | Value |
|---|---|---|
| Initial position of missile | $(x_M, z_M)$ | (0, 0) m |
| Initial flight-path angle of missile | $\eta$ | 0 deg |
| Velocity magnitude of missile | $V_M$ | 600 m/s |
| Initial position of target | $(x_T, z_T)$ | (10000, 0) m |
| Initial flight-path angle of target | $\beta$ | 170 deg |
| Target velocity magnitude | $V_T$ | 300 m/s |
| Target acceleration | $a_T$ | 9 g, S-type maneuver |

The numerical simulation results of Algorithm 3 are compared with the adaptive dynamic surface guidance (ADSG) law and linear quadratic differential game (LQDG) introduced in [36] and [37]. Numerical simulation results of missile target engagement process are shown in Figs. 7-10. Fig. 7 shows the missile and target trajectories, the miss distance is 0.082 m, 0.200 and 1.122 m by using α-PI algorithm, ADSG and LQDG respectively. α-PI algorithm and ADSG can achieve a better guidance accuracy. The acceleration instruction histories of the three guidance algorithms are demonstrated in Fig. 8. In the initial phase, the target is non-maneuvering, and after 1.5 s, the target actuates a S-type maneuver with acceleration of 9 g, i.e., the maneuvering acceleration is 9 times that of gravity. In the initial, the acceleration instruction of missile guided by α-PI algorithm is driven by the white noise. Compared with LQDG, α-PI algorithm and ASDG have a lower demand for missile acceleration. Because the α-PI algorithm solves the approximate solution of the nonlinear differential game guidance problem, which has the advantages of low demand of acceleration and high guidance accuracy. Furthermore, compared to ASDG, there is no chattering phenomenon in the acceleration instruction of α-PI algorithm. The actor NN weights learned during the online learning guidance process are shown in Fig. 9. Fig. 10 shows the number of iterations per calculation cycle, as can be seen that the maximum iteration number is 46 at $t = 2.0$ s.

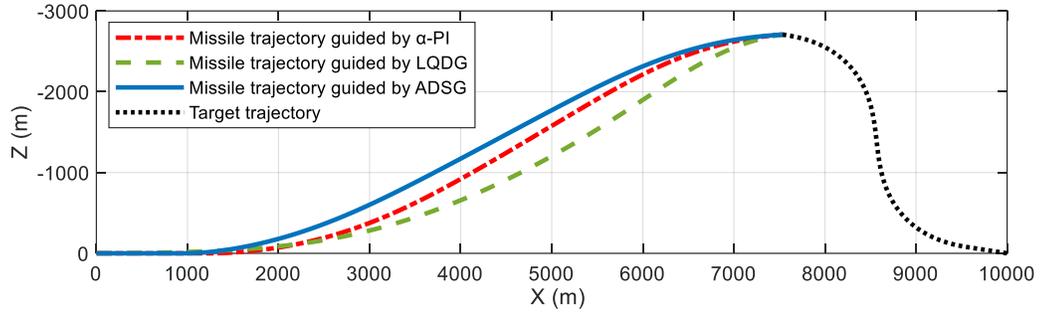

Fig. 7 Missile and target trajectories.

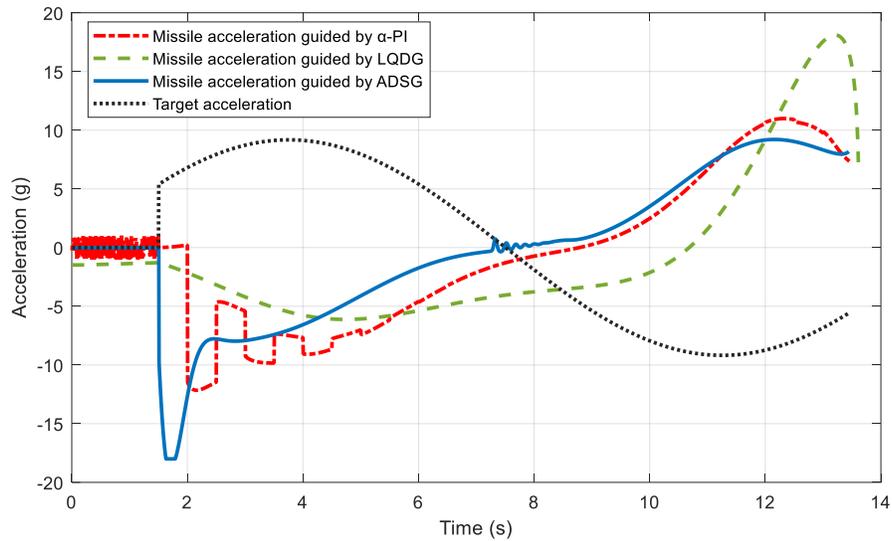

Fig. 8 Missile acceleration instruction for intercepting S-type maneuvering target.

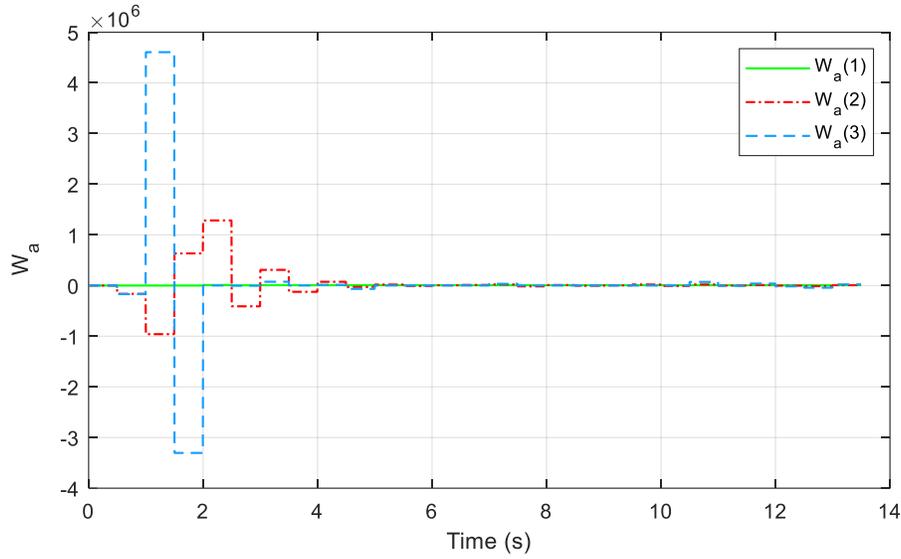

**Fig. 9   Actor NN weights learned during the online learning guidance process.**

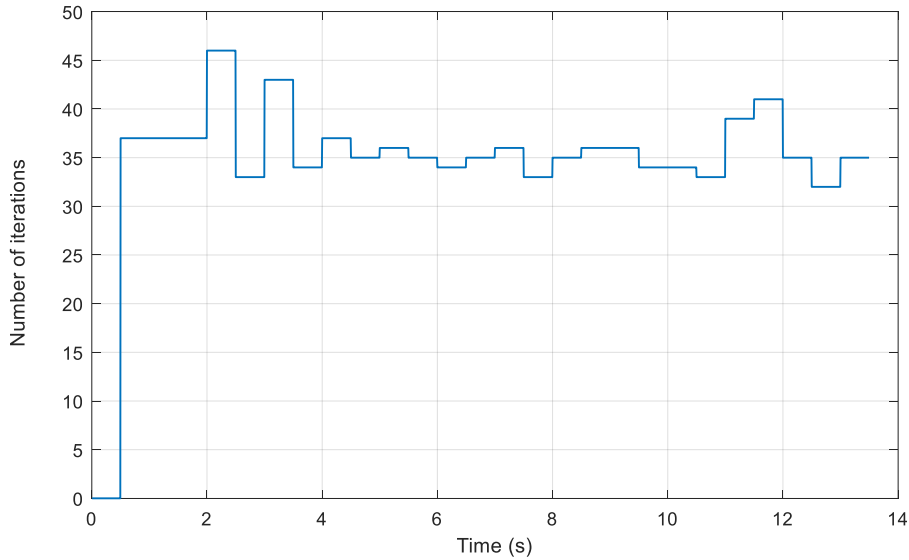

**Fig. 10   Number of iterations per calculation cycle.**

## VI. Conclusion

Till present, the development of reinforcement learning algorithms and theories for $H_\infty$ control design problem of continuous-time nonlinear system is still an open issue. Most integral reinforcement learning algorithms were based on Newton iterative method. We proposed a novel $\alpha$-PI algorithm which is based on the damped Newton iterative method in this paper. Damped Newton iterative method makes the convergence slower than Newton's method, but might be globally convergent. Thus $\alpha$-PI algorithm has better convergence theoretically. The generalized Bellman equation was obtained first, then, on-policy and off-policy $\alpha$-PI algorithms were derived, respectively. The on-policy $\alpha$-PI algorithm can be implemented online for pursuing the solution of $H_\infty$ control problem without using information about the system internal dynamics. The off-policy version of $\alpha$-PI approach is a model-free algorithm for solving $H_\infty$ control online. The implementation schemes based on neural network for these two $\alpha$-PI algorithms were proposed. And finally, the effectiveness of the off-policy $\alpha$-PI approach proposed is verified through two computer simulation examples, one is for the general nonlinear system $H_\infty$ control problem, and the other is missile interception simulation. In the Banach space, how to select the step size $\alpha$ in the $\alpha$-PI algorithm to obtain as global as possible convergence and fast convergence will be the focus of future research.